\documentclass[review]{elsarticle}

\usepackage{lineno,hyperref, amsmath, subcaption}
\modulolinenumbers[5]

\journal{Journal of \LaTeX\ Templates}

\begin{document}

\begin{frontmatter}

\title{Spin polarization induced by an electric field   in the presence of weak localization effects}

\author{Daniele Guerci$^{a}$}
\author{Juan Borge$^{b,a}$}
\author{Roberto Raimondi$^{a}$
\corref{mycorrespondingauthor}}
\cortext[mycorrespondingauthor]{Corresponding author}
\ead{roberto.raimondi@uniroma3.it}
\address{$^{a}$ Dipartimento di Matematica e Fisica, Universit\`a Roma Tre, Via della Vasca Navale 84, 00146 Roma, Italy}
\address{$^{b}$  Nano-Bio Spectroscopy group, Departamento de F\`isica de Materiales,
Universidad del Pa\`is Vasco UPV/EHU, E-20018 San Sebasti\`an, Spain}




\begin{abstract}
We evaluate the  spin polarization (Edelstein or inverse spin galvanic effect)
 and the spin Hall current induced by an applied electric field by including the weak localization corrections for a two-dimensional electron gas. We show that  the weak localization effects yield
logarithmic corrections to both the spin polarization conductivity relating the spin polarization and the electric field  and to the spin Hall angle relating the spin and charge currents. The renormalization of both the spin polarization conductivity  and the spin Hall angle combine to produce  a zero correction to the total spin Hall conductivity as required by an exact identity. Suggestions for the experimental observation of the effect are given.
\end{abstract}

\begin{keyword}
%\texttt{elsarticle.cls}\sep \LaTeX\sep Elsevier \sep template
%\MSC[2010] 00-01\sep  99-00
Spin-orbit coupling, weak localization, 2DEG, spintronics
\end{keyword}

\end{frontmatter}

\linenumbers

\section{Introduction}

Weak localization (WL) is the result of quantum interference corrections to the semiclassical theory of transport\cite{bergman1984,lee1985}. It manifests itself in good conductors as a negative or positive correction to the electrical conductivity depending on the  symmetry properties of the system. The functional form varies with the effective dimensionality of the sample,
behaving as a square root of temperature in three dimensions and logarithmically in two dimensions\cite{gorkov1979}. In the latter case, the resummation of the logarithmic correction via the renormalization group leads eventually to the Anderson localization transition in $d=2+\epsilon$ dimensions\cite{abrahams1979,efetov1980}. In the presence of spin-orbit coupling (SOC), the correction is positive and hence manifests as  an antilocalizing behavior\cite{hikami1980}.  SOC affects WL because it yields a finite spin relaxation time, which introduces a cutoff in the logarithmic singularity associated with the so-called triplet channel of the particle-particle ladder, known as the Cooperon.
Since the singlet and the triplet channels contribute to WL with opposite signs, the elimination of the triplet leaves the singlet alone, which then produces the antilocalizing behavior. In metallic conductors and doped semiconductors SOC was traditionally attributed to the electric field of impurities, which do not affect the nature of the electron eigenstates. In the last two decades, however, the two-dimensional electron gas (2DEG) has become one of the most analyzed model systems for electrical transport, due to the possibility of realizing it in semiconducting systems,
and more recently at metallic\cite{RojasSanchez2013} and oxides\cite{Caviglia10} interfaces.
The realization of the 2DEG leads to the breaking of inversion symmetry with respect to the axis, say the z axis, perpendicular to the 2DEG plane, say the x and y plane. In these circumstances, in the presence of a finite spin-orbit interaction, Bychkov and Rashba have proposed a model Hamiltonian\cite{Rashba84}, which, besides the standard effective-mass kinetic energy term, contains a spin-orbit coupling term linear in momentum
\begin{equation}
\label{Rashba_Ham}
H=\frac{p^2}{2m}+\alpha (\sigma^x p_y -\sigma^y p_x),
\end{equation}
where ${\bf p}=(p_x,p_y)$ is the vector of the components of the momentum operator, $m$ is the effective mass and $\alpha$ a SOC constant with $\sigma^x$ and $\sigma^y$ the standard Pauli matrices. The Rashba Hamiltonian Eq.(\ref{Rashba_Ham}) has been extensively studied over the last twenty years, especially   aiming at the development of new spintronic functionalities\cite{Awschalom2007}. In this respect the spin Hall effect (SHE)\cite{Dyakonov71,Hirsch99,Zhang00,Murak04,Sinova04} and the current-induced spin polarization effect\cite{Edelstein1990,Lyanda-Geller89}
(known also as the Edelstein or inverse spin-galvanic effect) have been the focus of an intensive dedicated research.
These effects, whose precise definition will be given later on, manifest due to the coupling of charge and spin degrees of freedom and hence introduce, besides the standard electrical conductivity, new transport parameters.
These are defined as the linear coefficients relating the spin polarization and the spin current to the applied electric field
\begin{equation}
\label{definitions}
\langle \langle s^y\rangle \rangle= \sigma^{EC} E_x, \ \ \
\langle \langle j^z_y\rangle\rangle = \sigma^{SHC} E_x,
\end{equation}
where the double brackets indicate the quantum and statistical average. $\sigma^{EC}$ and  $\sigma^{SHC} $ are  referred to as the
 spin polarization  or Edelstein  and 
  the spin Hall conductivities, respectively.  As for the electrical conductivity, these transport parameters can be studied with the well-known impurity technique.
One advantage of this technique\cite{Abrikosov:QFT}, based on standard diagrammatic perturbation theory, is the appearance of the semiclassical Drude-Boltzmann theory of transport at the leading approximation in an expansion of the small parameter $\hbar /(\epsilon_F \tau )$, where $\epsilon_F$ and $\tau$ are the Fermi energy and the elastic scattering time,
the only two parameters characterizing a disordered Fermi gas. In such an expansion, WL arises in the next-to-leading approximation in the expansion in $\hbar /(\epsilon_F \tau )$.

WL effects in the presence of the Rashba SOC described by Eq.(\ref{Rashba_Ham}) have been analyzed by several authors,  most of the attention having been focused on the electrical conductivity only\cite{Iordanskii_1994,Edelstein95,Skvortsov_98,Lyanda98,Wenk_10,Araki_2014}. It is the aim of the present work
to extend this analysis to the other transport parameters mentioned above, whose experimental study has developed considerably in the last few years\cite{Jungwirth2012,RojasSanchez2013}.   We find that  $\sigma^{EC}$ and the spin Hall angle  $\gamma_{SH}=e \sigma^{SHC}_{drift}/\sigma_0$ acquire logarithmic corrections which can be absorbed in terms of the renormalization of the scattering time appearing in the electrical conductivity
$\sigma_0$. 
We emphasize that $\sigma_{drift}^{SHC}$ is not the full spin conductivity $\sigma^{SHC}$ which would be measured in an experiment\cite{Shen2013}. 
As  will be shown in the next Section,
$\sigma^{SHC}$ can be expressed in terms of $\sigma^{EC}$ and  $\sigma^{SHC}_{drift}$.
The renormalizations of both $\sigma^{EC}$  and $\gamma_{SH}$  compensate in such a way  that  $\sigma^{SHC}$ has no correction as expected on general arguments\cite{Dimitrova05}.
% and hence signal an independent renormalization. These corrections arise from the interplay of singular backscattering of WL and Rashba SOC.
%
%NOTE FOR US. LATER WE MUST DEVELOP MORE %PHYSICAL INTUITION HERE.

The plan of the paper is as follows. In the next Section we introduce the disordered Rashba model and
review the theory of $\sigma^{EC}$ and $\sigma^{SHC}$ to the leading order in the parameter $\hbar /(\epsilon_F \tau )$ within the impurity technique. This is necessary to prepare the ground for the following Sections. Section III deals with the WL localization corrections in the presence of the Rashba SOC. The evaluation of the electrical conductivity is reviewed as an example. Section IV presents  the calculation of the WL corrections to $\sigma^{EC}$ and $\sigma^{SHC}$.
Section V provides a discussion of the results obtained,
whereas technical points of the calculations are
given in the appendices at the end of the paper.
From now on, if not otherwise specified, we will work in natural units $\hbar=c=1$.

\section{The disordered Rashba two-dimensional electron gas at leading order in $1/(\epsilon_F \tau)$}
In the presence of scattering from impurities,
the 2DEG Hamiltonian of Eq.(\ref{Rashba_Ham}) acquires  an additional random potential term  $U({\bf r})$ defined by
the averages
\begin{equation}
\label{disorder}
\langle U({\bf r})\rangle =0, \ \
\langle U({\bf r} )U({\bf r} ')\rangle = \frac{1}{2\pi N_0 \tau}\delta ({\bf r} -{\bf r}')
\end{equation}
where ${\bf r}=(x,y)$ and ${\bf r'}=(x',y')$ are the coordinate operators,
$N_0=m/2\pi$  the two-dimensional density of states and $\tau$ the elastic scattering time.
At leading order in the expansion parameter $1/(\epsilon_F \tau)$, the selfenergy is given by the
selfconsistent Born approximation
\begin{equation}
\label{self1}
\Sigma^{R,A}( {\bf p},\epsilon)=\frac{1}{2\pi N_0\tau}\sum_{{\bf p'}}G^{R,A}({\bf p'},\epsilon),
\end{equation}
where $G^{R,A}$ denotes the retarded and advanced Green functions. As discussed in \cite{Raimondi02,Raimondi05}, in the presence of
 Rashba SOC the Green function has a nontrivial structure in spin space, whereas the selfenergy remains diagonal, $\Sigma=\Sigma^0\sigma^0$, $G=G^0\sigma^0+G^1\sigma^1+G^2\sigma^2$.  Explicitly we have:
\begin{eqnarray}
\label{G1}
\nonumber
G^0&=&\frac{1}{2}(G_++G_-)\\
\nonumber
G^1&=&\frac{{\hat p}_y}{2}(G_+-G_-)\\
\nonumber
G^2&=&-\frac{{\hat p}_x}{2}(G_+-G_-)\\
G_{\pm}&=&\left(\epsilon+\mu-p^2/2m\mp \alpha p-\Sigma^0\right)^{-1},
\end{eqnarray}
with
\begin{equation}
\label{self2}
(\Sigma^0)^{R,A}=\mp\frac{i}{2\tau}.
\end{equation}
The Edelstein (EC) and spin Hall (SHC) conductivities are defined in terms of the spin polarization and spin Hall current induced by an applied electric field taken along the x axis for definiteness's sake $E_x=-\partial_t A_x$. The corresponding Kubo formulae are
\begin{equation}
\label{E2_0}
\sigma^{EC}=\lim_{\omega\rightarrow 0} \frac{{\rm Im}\langle\langle s^y;j_x\rangle\rangle}{\omega},
\end{equation}
and
\begin{equation}
\label{SHC2_0}
\sigma^{SHC}=\lim_{\omega\rightarrow 0} \frac{{\rm Im}\langle\langle j^z_y;j_x\rangle\rangle}{\omega},
\end{equation}
where the bare vertices $s^y=\sigma^y /2$,  $j^z_y=\sigma^z p_y /2m$ and $j_x=-e\hat v_x$, $\hat v_x ({\bf p})=p_x/m-\alpha \sigma^y$ denote the operators for spin polarization, spin current and charge current, respectively.
%The double angular brackets indicate the quantum and statistical averages.
The evaluation of the response functions (\ref{E2_0}-\ref{SHC2_0}) involves the standard bubble diagrams  of the Green function lines obtained by the selfconsistent Born approximation (\ref{self1}) decorated by the insertion of the impurity ladder. This corresponds to the inclusion of the so-called vertex corrections, which lead to renormalized vertices\cite{Raimondi02}.

The  expression (\ref{E2_0}) for the EC becomes
\begin{equation}
\label{8}
\sigma^{EC}=-\frac{e}{2\pi}\sum_{\bf{p}}\text{Tr}\Big[S^{y}G^R_{\bf{p}}\hat{v}_{x}({\bf{p}})G^A_{\bf{p}}\Big],
\end{equation}
where the vertex renormalization can be attributed either to the left spin vertex or to the right current vertex and we have dropped the  dependence on the frequency argument of the Green function. In the former case, by
using the renormalized spin vertex indicated by a capital letter
$S^y=((1+x^2)/x^2)\sigma^{y}\equiv (\tau_{DP}/2\tau)\sigma^y=(\tau_{DP}/\tau)s^y$, one obtains\cite{Edelstein1990}
\begin{equation}
\label{E6}
\sigma^{EC}_0=-e\alpha N_0\tau,
\end{equation}
where the subscript $0$ in $\sigma^{EC}_0$ indicates the lowest order in the parameter $1/(\epsilon_F \tau)$. We have defined the parameter $x=2\alpha p_F \tau$ and introduced the D'yakonov-Perel relaxation time $\tau_{DP}=2\tau ((1+x^2)/x^2)$, $p_F$ being the Fermi momentum in the absence of the Rashba SOC.
The model has two small parameters $1/(\epsilon_F \tau)$ and $\alpha /v_F$ with $v_F=p_F/m$, in terms of which the above paramater $x=4(\alpha /v_F)(\epsilon_F \tau)$ can be expressed.

Similarly, for the SHC one has the expression
\begin{equation}
\sigma^{SHC}=-\frac{e}{2\pi} \sum_{\bf{p}} \text{Tr}\Big[J^{z}_{y}({\bf{p}})G^R_{\bf{p}}\hat{v}_{x}({\bf{p}})G^A_{\bf{p}}\Big].
\end{equation}
By using the renormalized spin current vertex $J^{z}_{y}=j_y^z+(v_F/2x)\sigma^y=j_y^z+(v_F/2) (x/(1+x^2))S^y$ \cite{Tse_IntrinSH_PRB06}, one gets
\begin{eqnarray}
\nonumber
\sigma^{SHC}&=&-\frac{e}{2\pi}\sum_{\bf{p}}\left(\text{Tr}\Big[j^z_y G^R_{\bf{p}}\hat{v}_{x}({\bf{p}})G^A_{\bf{p}}\Big]+\frac{v_F}{2}\frac{x}{1+x^2}\text{Tr}\Big[S^y G^R_{\bf{p}}\hat{v}_{x}({\bf{p}})G^A_{\bf{p}}\Big]\right)\\
&=&\sigma^{SHC}_{drift}+\frac{v_F}{2}\frac{x}{1+x^2}\sigma_0^{EC}\nonumber\\
&=&\frac{e}{8\pi}\frac{x^2}{1+x^2}+\frac{v_F}{2}\frac{x}{1+x^2}\sigma_0^{EC},\label{SHC5}
\end{eqnarray}
where we indicated by $\sigma^{SHC}_{drift}$ the SHC in the absence of vertex corrections as first computed in Ref. \cite{Sinova04}.  It is then not difficult to see that the insertion of the result (\ref{E6}) into (\ref{SHC5}) gives a vanishing SHC. This latter result
is actually expected
following the argument derived by Dimitrova \cite{Dimitrova05}. The commutation relation allows us to write
\begin{equation}
\label{Di2}
\frac{{\rm d }s^y}{{\rm d}t}=-2m\alpha j_y^z,
\end{equation}
which, in stationary circumstances, implies $\langle\langle j^z_y\rangle\rangle=0$.

The expression (\ref{SHC5})  shows that the vertex corrections for the SHC are associated to the EC. This connection between the two effects acquires a more transparent meaning by adopting the SU(2) gauge-field description of the SOC\cite{Raimondi_AnnPhys12}. In such a picture, the SOC is expressed
 in terms of a non-Abelian gauge field
${\boldsymbol {\mathcal A}}={\mathcal A}^a\sigma^a/2$, with ${\mathcal A}^x_y= 2m\alpha$ and ${\mathcal A}^y_x=-2m\alpha$ \cite{Tokatly_Color_PRL08,Gorini10}. The first consequence of resorting to this language is the appearance of an SU(2) magnetic field
${\cal B}^z_z=-(2m\alpha)^2$, which arises from the non-commuting components of the Bychkov-Rashba vector potential. As in the normal Hall effect we have a spin Hall drift component of the spin current which can be described  as (assuming $x\ll 1$)
\begin{equation}
\label{SHDRIFT}
\langle\langle[j^z_y]_{drift}\rangle\rangle=\sigma^{SHC}_{drift} E_x,
\end{equation}
with $\sigma^{SHC}_{drift}=(e/8\pi) x^2$.
In addition to the drift current, there is also a ``diffusion current'' due to spin precession around the effective spin-orbit field. Within the SU(2) formalism this current arises from the replacement of the ordinary derivative with the SU(2) covariant derivative in the expression for the diffusion current.  The SU(2) covariant derivative, due to the gauge field, is
\begin{equation}
\nabla_j {\cal O} = \partial_j {\cal O} + i\left[{\mathcal A}_j,{\cal O}\right],
\end{equation}
with ${\cal O}$ a given quantity being acted upon.  The normal derivative, $\partial_j$, along a given axis $j$
is shifted by the commutator with the gauge field component along that same axis.
As a result of the replacement $\partial \to \nabla$ diffusion-like terms,
normally proportional to spin density gradients,
arise even in uniform conditions and the diffusion contribution to the spin current turns out to be
\begin{equation}
\label{DIFFUSION}
\langle\langle[j^z_y]_{diff}\rangle\rangle= \frac{2m\alpha}{\hbar} D \langle\langle s^y\rangle\rangle,
\end{equation}
where $D=v_F^2\tau /2$ is the diffusion coefficient.
In the diffusive regime the full spin current $j^z_y$ can thus be expressed as
\begin{equation}
\label{discussion_1}
\langle\langle j^z_y\rangle\rangle=\frac{2m\alpha}{\hbar} D\langle\langle s^y\rangle\rangle +\frac{\gamma_{SH}}{e}\langle\langle j_x\rangle\rangle,
\end{equation}
with  the spin Hall angle given by $\gamma_{SH}=e \sigma^{SHC}_{drift}/\sigma_0=m\alpha^2\tau$  and $\langle \langle j_x\rangle\rangle =\sigma E_x$.
Relation Eq.(\ref{discussion_1}), which implies   the diffusive limit, coincides with the expression (\ref{SHC5}) when $x\ll 1$, i.e. when the spin-orbit induced spin splitting is much smaller than the  disorder broadening of the levels.

\section{The weak-localization correction to the electrical conductivity in the presence of Rashba spin-orbit coupling}
In this Section we consider the WL corrections, which arise to first order in $1/(\epsilon_F \tau )$. In next Subsection we review the evaluation of the Cooperon propagator, whereas in the following one we apply it to the case of the  electrical conductivity. The spin polarization and spin current response are considered in the next Section.
\subsection{The structure of the Cooperon in spin space}
 \begin{figure}
\begin{center}
\includegraphics[width=\textwidth]{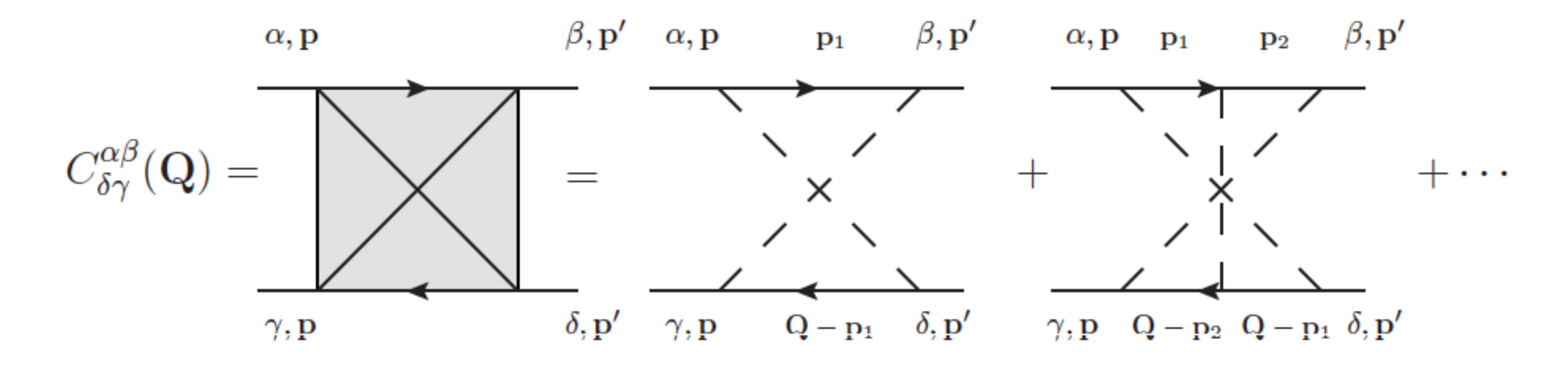}
\caption{Crossed ladder, also known as Cooperon. The indices $\alpha$ and $\delta$ are the incoming spin indices, while $\beta$ and $\gamma$ are the outgoing ones.
Furthermore, the momentum ${\bf Q}$ is the sum of the initial and final external momenta in each Green function line. }
\label{fig:Cooperon}
\end{center}
\end{figure}
In the absence of SOC, the Cooperon obeys the diffusion equation, which in momentum space reads
\begin{equation}
\label{coop_1}
D Q^2 C({\bf Q})=\frac{1}{2\pi N_0\tau^2},
\end{equation}
where ${\bf Q}={\bf p}+{\bf p'}$ is the sum of the initial and final momenta in each Green function line, as shown in Fig. \ref{fig:Cooperon}.
 We remind that  the particle-hole series of the maximally crossed diagrams can be transformed in the particle-particle ladder by reversing one of the two Green function lines and the associated momentum.
The momentum  ${\bf Q}$ corresponds then to the total momentum of the particle-particle pair of the Cooperon propagator.
The SU(2) gauge-field point of view suggests that one can obtain the equation for the Cooperon by a minimal substitution procedure on the momenta of both Green function lines, i.e. ${\bf p}\rightarrow {\bf p}+{\bf \cal A}$ and ${\bf p'}\rightarrow {\bf p'}+{\bf \cal A}$ , where ${\bf \cal A}$ is the SU(2) spin-dependent vector potentials introduced in the previous Section. The equation for the Cooperon becomes then
\begin{equation}
\label{coop_2}
D \sum_{i=1}^2 (Q_i+{\cal A}_i^a \hat S^a)^2 C({\bf Q})= \frac{1}{2\pi N_0\tau^2},
\end{equation}
where $\hat S^a=\sigma_1^a/2\oplus\sigma_2^a/2$ is the total spin of the two particles ($\sigma^a_{1,2}$ refers to the upper/lower Green function line).  The Cooperon acquires a matrix structure acting in the four-dimensional Hilbert space resulting from the combination of the two  spin one-half spaces.
It is convenient to use the triplet-singlet basis $\big\{
|\uparrow\uparrow\rangle, (|\uparrow\downarrow\rangle+|\downarrow\uparrow\rangle)/\sqrt{2}, |\downarrow\downarrow\rangle, (|\uparrow\downarrow\rangle-|\downarrow\uparrow\rangle)/\sqrt{2}\big\}$, in terms of which the Cooperon becomes block-diagonal. Since for the singlet the total spin of the pair is zero $S=0$, the one-dimensional corresponding block has the same form as in the absence of SOC. The $S=1$ triplet three-dimensional block can be obtained from Eq.(\ref{coop_2}) by expanding and inverting the square in the left hand side. One sees that, besides the standard diffusive terms going as $Q^2$, which are diagonal in spin space, linear-in-$Q$ terms arise when considering the double product. Finally, the terms arising from the square of the spin-dependent vector potential  give rise to $Q$-independent terms. The latter are also  diagonal in spin space and yield a finite spin relaxation rate. The  full expression for the Cooperon  is obtained by inverting $C^{-1}({\bf Q})$
\cite{Iordanskii_1994,Edelstein95,Skvortsov_98,Lyanda98, Wenk_10}
\small
    \begin{equation}
\label{KW1}
C^{-1}({\bf Q})=2\pi N_{0}\tau\,\left(\begin{array}{cccc}D\tau Q^{2}+\tau/\tau_{DP} & ie^{-i\theta}x\,Q\sqrt{D\tau} & 0 & 0 \\-ie^{i\theta}x\,Q\sqrt{D\tau} & D\tau Q^{2}+2\tau/\tau_{DP} & ie^{-i\theta}x\,Q\sqrt{D\tau} & 0 \\0 & -ie^{i\theta}x\,Q\sqrt{D\tau} & D\tau Q^{2}+\tau/\tau_{DP} & 0 \\0 & 0 & 0 & D\tau Q^2\end{array}\right),
\end{equation}
\normalsize
where $\theta$ is the angle between ${\bf Q}$ and the $x$-axis.
According to the matrix form (\ref{KW1}), the singlet state decouples from the triplet states and remains gapless, whereas the triplet sector acquires a gap proportional to $\tau/\tau_{DP}$. Furthermore, the linear-in-$Q$ terms provide the mixing between the different triplet channels.
These results are due to the D'yakonov-Perel spin relaxation mechanism, which mixes the triplet states and kills their singular contribution.

We notice that the previous expression (\ref{KW1}) is valid in the regime $x\ll1$. 
In \ref{CRSOC} we derive the exact form of the Cooperon equation (\ref{CTS}), whose solution tends to expression (\ref{KW1}) as $x\ll1$.
   However, the approximate solution (\ref{KW1})  possesses all the physical features due to the presence of Rashba SOC and simplifies the calculations of the quantum corrections.
   Because of that, in the rest of this work we will use the Cooperon expression (\ref{KW1}) to calculate the weak localization corrections.

\subsection{The electrical conductivity to order $1/(\epsilon_F \tau )$}

\begin{figure}
        \centering
        \begin{subfigure}[b]{0.5\textwidth}
                \includegraphics[width=\textwidth]{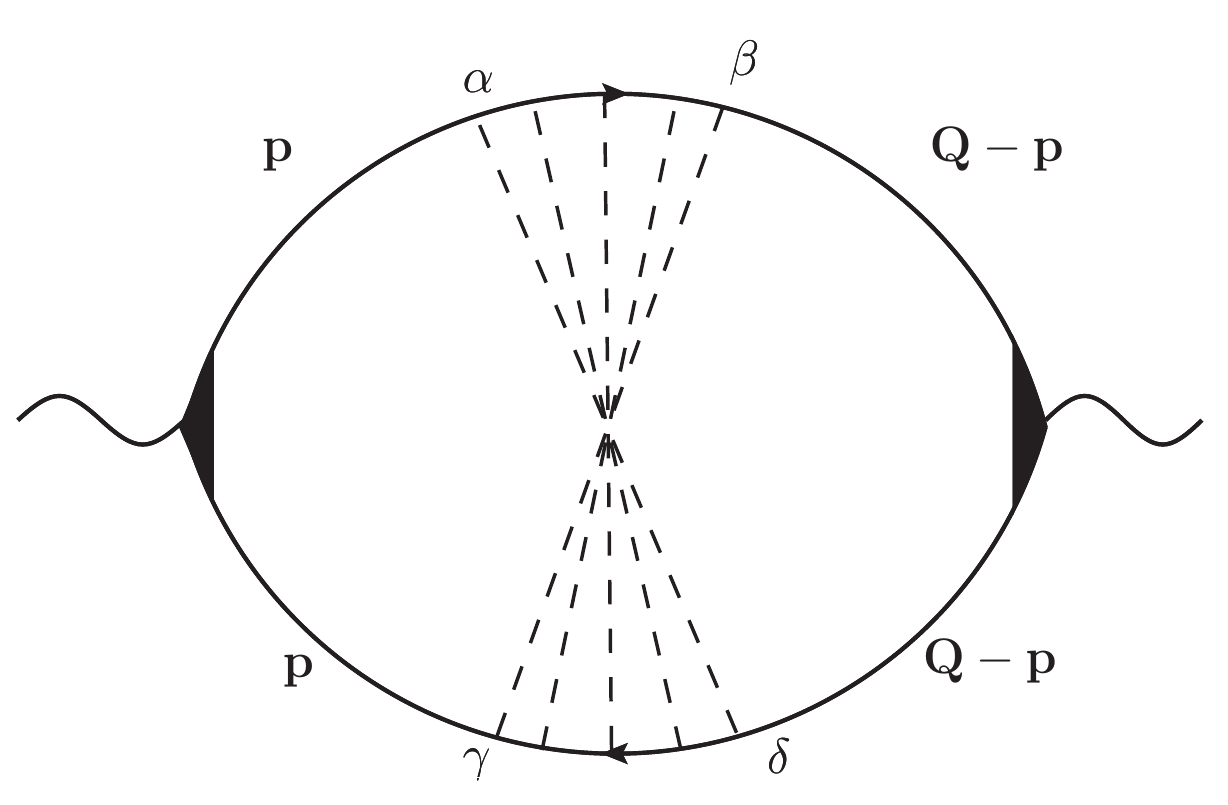}
		\caption{}
		\label{bubble}
        \end{subfigure}%
        ~ %add desired spacing between images, e. g. ~, \quad, \qquad, \hfill etc.
        \begin{subfigure}[b]{0.5\textwidth}
                \includegraphics[width=\textwidth]{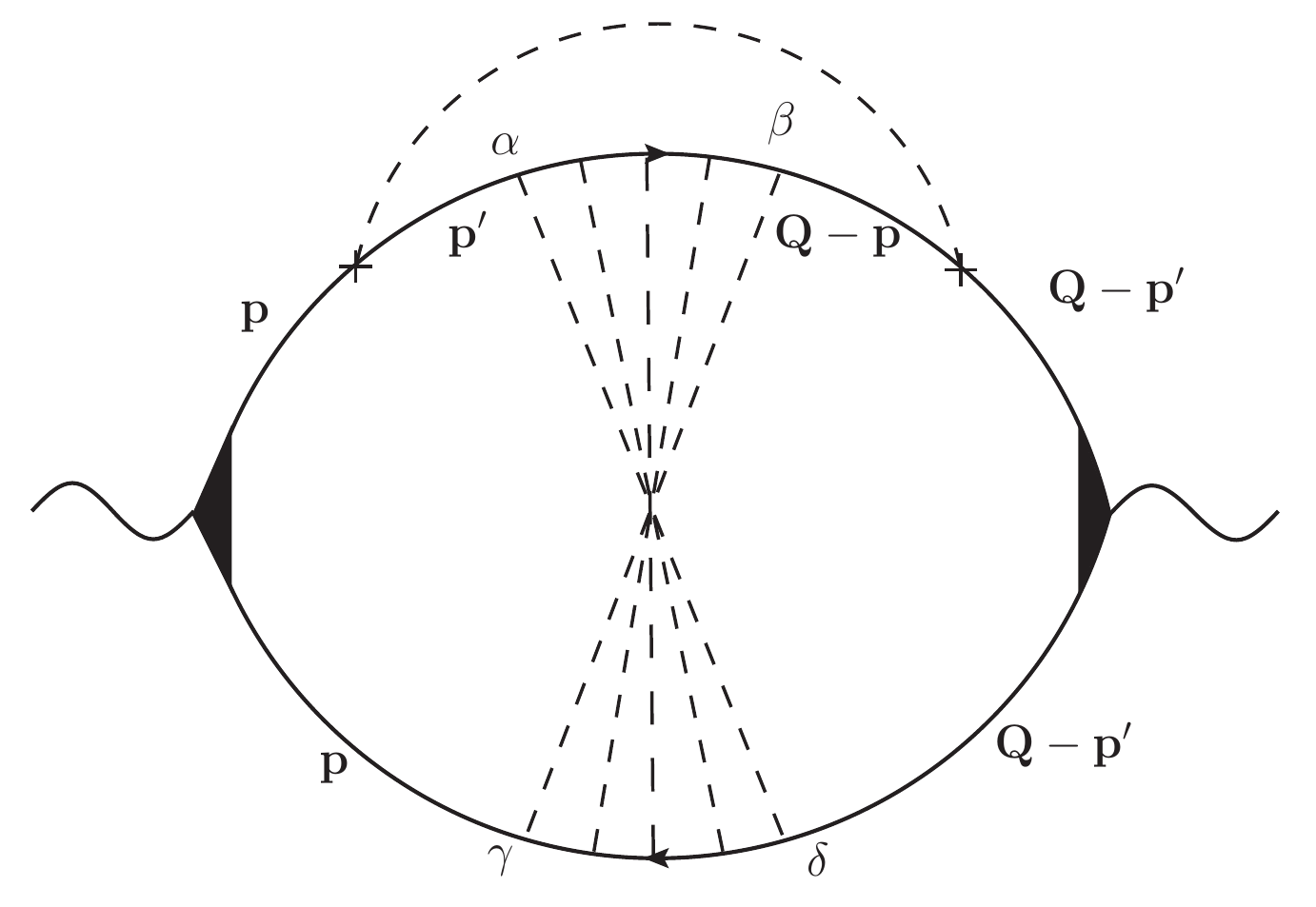}
                \caption{}
                \label{up}
        \end{subfigure}
        ~ %add desired spacing between images, e. g. ~, \quad, \qquad, \hfill etc.
        \begin{subfigure}[b]{0.5\textwidth}
                \includegraphics[width=\textwidth]{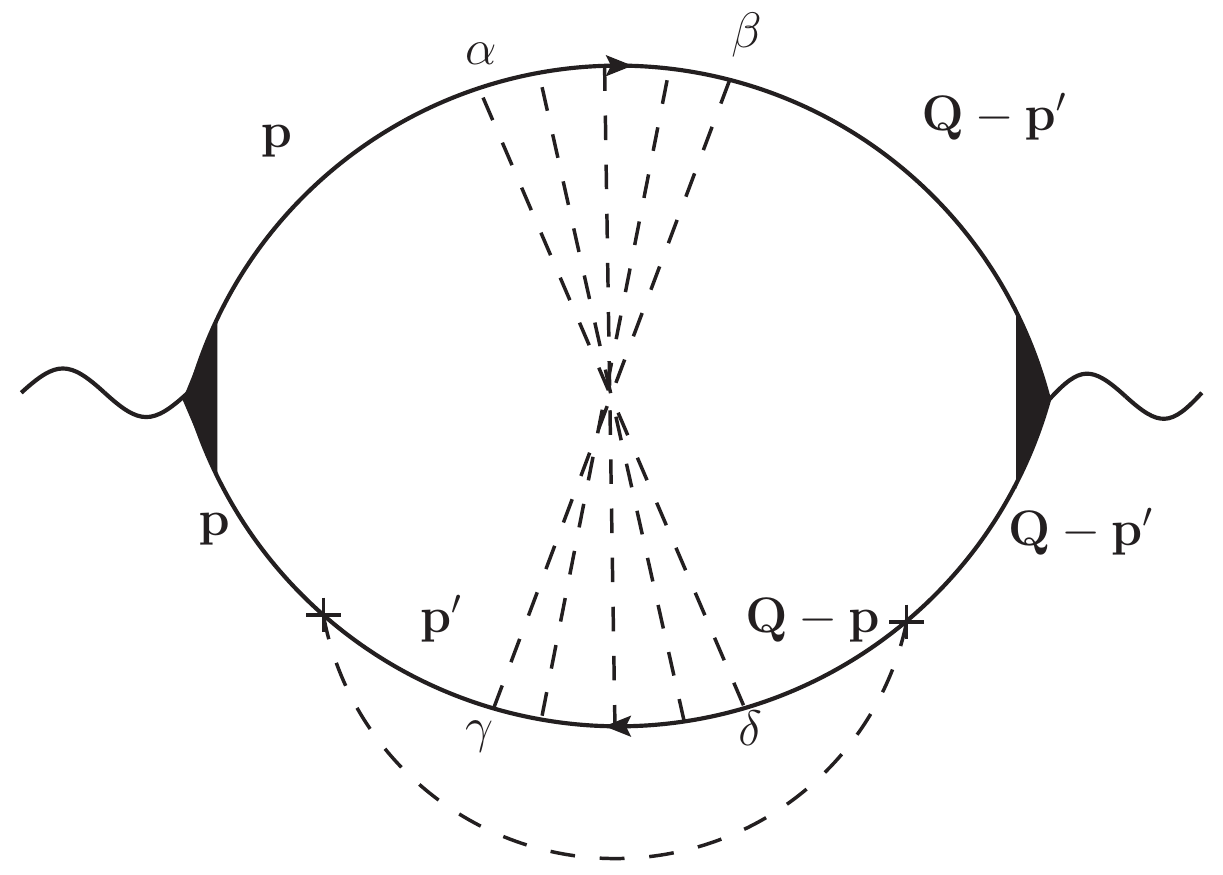}
                \caption{}
                \label{down}
        \end{subfigure}
        \caption{Maximally crossed diagrams giving the weak localization corrections to the transport coefficients.
     In each diagram the ladder summation takes into account the renormalization of both vertices of the bubble. The crossed lines indicate the crossed ladder, also known as Cooperon.
     In diagram (a) the integration over the momentum ${\bf p}$ of four Green functions yields the so-called Hikami box, whose full expression is obtained by adding the integrations over ${\bf p}$ and ${\bf p}^{\prime}$ of diagrams (b) and (c). }
     \label{Diagrams}
\end{figure}

First we note that the resummation of the maximally crossed diagrams for the evaluation of the electrical conductivity  gives rise to three diagrams, depicted in Fig. \ref{Diagrams}. The following discussion is quite general and applies also to the case of EC and SHC to be discussed in the next Section.
%To analyze these latter cases it is sufficient to replace
 %the left current vertex with the renormalized spin current or the renormalized spin density vertex.

The weak localization correction to the electrical conductivity reads
\begin{equation}
\label{QCDC}
\delta\sigma=\delta\sigma_{(a)}+\delta\sigma_{(b)}+\delta\sigma_{(c)}\,,
\end{equation}
where the subscript refers to each diagram in Fig. \ref{Diagrams}.
As  can be understood by looking at Fig. \ref{bubble}, the first diagram gives the following contribution
\begin{equation}
\label{WLCa1}
\delta\sigma_{(a)}=\frac{1}{2\pi}\sum_{\rho\mu\nu\lambda}\text{Tr}(\sigma^{\mu}\sigma^{\lambda}\sigma^{\nu}\sigma^{\rho})\sum_{{\bf Q}}C^{\mu\nu}({\bf Q})S^{\rho\lambda}_{(a)}({\bf Q})\,,
\end{equation}
with
\begin{equation}
\label{Coop}
C^{\mu\nu}({\bf Q})=\frac{1}{4}\text{Tr}(L^{\mu\nu} \hat{C}({\bf Q}))
\end{equation}
where $L^{\mu\nu}=\sigma^{\mu}\otimes\sigma^{\nu}$ is the direct product basis with $\nu,\mu=0,1,2,3$, and
\begin{equation}
\label{WFa0}
S^{\rho\lambda}_{(a)}({\bf Q})=\sum_{{\bf p}}\frac{1}{2}\text{Tr}\Big[G^{A}_{{\bf p}}(-e)\frac{p_{x}}{m}G^{R}_{{\bf p}}\sigma^{\rho}\Big]\frac{1}{2}\text{Tr}\Big[G^{R}_{{\bf Q}-{\bf p}}(-e)\frac{Q_{x}-p_{x}}{m}G^{A}_{{\bf Q}-{\bf p}}\sigma^{\lambda}\Big]\,.
\end{equation}
Notice that the renormalized charge current reads
$J_x=(-e)p_x/m=(-e)(\hat v_x+\alpha\sigma^y)$\cite{Raimondi02}.
In order to evaluate Eq.(\ref{WFa0}) we make three approximations: first, we cut off the sum over ${\bf Q}$; second, we expand $C({\bf Q})$ for small ${\bf Q}$; third, we neglect the weak ${\bf Q}$ dependence of $S^{\rho\lambda}_{(a)}({\bf Q})$. The natural upper cut-off of the sum over ${\bf Q}$ is provided by the inverse of the mean free path, i.e. the length where the diffusive behaviour sets in.
On the other hand, the lower cut-off of the sum is given by the inverse of  the maximal size of a loop allowed to contribute to the coherent backscattering process. Then, for a system of linear size $L$ the lower cut-off is $1/L$.
For small ${\bf Q}$ the Cooperon is strongly depending on ${\bf Q}$ and gives a divergent contribution which is not modified by the weak ${\bf Q}$ dependence of the weight factors $S^{\rho\lambda}_{(a)}({\bf Q})$.
  Eventually, we arrive at the following expression for the conductivity correction given by the diagram in Fig. \ref{bubble}
   \begin{equation}
  \begin{split}
\label{a.1}
\delta\sigma_{(a)}=\frac{1}{2\pi}\sum_{\rho\mu\nu\lambda}\text{Tr}(\sigma^{\mu}\sigma^{\lambda}\sigma^{\nu}\sigma^{\rho})S^{\rho\lambda}_{(a)}C^{\mu\nu}\,,
\end{split}
\end{equation}
with the weight factors
\begin{equation}
\label{WFa1}
S^{\rho\lambda}_{(a)}=-\sum_{{\bf p}}\frac{1}{2}\text{Tr}\Big[G^{A}_{{\bf p}}(-e)\frac{p_{x}}{m}G^{R}_{{\bf p}}\sigma^{\rho}\Big]\frac{1}{2}\text{Tr}\Big[G^{R}_{-{\bf p}}(-e)\frac{p_{x}}{m}G^{A}_{-{\bf p}}\sigma^{\lambda}\Big]\,,
\end{equation}
and the ${\bf Q}$-integrated Cooperon matrix elements
\begin{equation}
\label{sum}
C^{\mu\nu}={\sum_{{\bf Q}}} C^{\mu\nu}({\bf Q})=\int_{0}^{2\pi}\frac{d\theta}{2\pi}\int_{1/L}^{1/l}\frac{Q\,dQ}{2\pi}C^{\mu\nu}({\bf Q})\,.
\end{equation}
   With the approximations introduced above, the contribution $\delta\sigma_{(b)}$, given by the diagram of Fig. \ref{up}, is
\begin{equation}
\label{b.1}
\delta\sigma_{(b)}=\frac{1}{2\pi}\sum_{\rho\mu\nu\lambda}\text{Tr}(\sigma^{\mu}\sigma^{\lambda}\sigma^{\nu}\sigma^{\rho})S^{\rho\lambda}_{(b)}C^{\mu\nu}\,,
\end{equation}
with the weight factors
\begin{equation}
\label{WFb}
S^{\rho\lambda}_{(b)}=-\frac{1}{2\pi N_{0}\tau}\sum_{{\bf p},{\bf p}^{\prime}}\frac{1}{2}\text{Tr}\Big[G^{A}_{{\bf p}}(-e)\frac{p_{x}}{m}G^{R}_{{\bf p}} G^{R}_{{\bf p}^{\prime}}\sigma^{\rho}\Big]\frac{1}{2}\text{Tr}\Big[G^{R}_{-{\bf p}} G^{R}_{-{\bf p}^{\prime}} (-e)\frac{p^{\prime}_{x}}{m}G^{A}_{-{\bf p}^{\prime}}\sigma^{\lambda}\Big]\,.
\end{equation}
Finally for
$\delta\sigma_{(c)}$ we find
 \begin{equation}
 \label{c}
\delta\sigma_{(c)}=\frac{1}{2\pi}\sum_{\rho\mu\nu\lambda}\text{Tr}(\sigma^{\mu}\sigma^{\lambda}\sigma^{\nu}\sigma^{\rho})S^{\rho\lambda}_{(c)}C^{\mu\nu}\,,
\end{equation}
where the weight factors can be expressed in terms of those of   $\delta\sigma_{(b)}$
\begin{equation}
\label{relationbc}
S^{\rho\lambda}_{(c)}=(S^{\rho\lambda}_{(b)})^{*}\,.
\end{equation}
%
%Overall, the weak localization correction to the conductivity is given by
%\begin{equation}
%\label{WLconductivity}
%\delta\sigma=\frac{1}{2\pi}\sum_{\rho\mu\nu\lambda}\text{Tr}(\sigma^{\mu}\sigma^{\lambda}\sigma^{\nu}\sigma^{\rho})(S^{\rho\lambda}_{(a)}+2\text{Re}S^{\rho\lambda}_{(b)})C^{\mu\nu}\,.
%\end{equation}
Upon integration over the directions of ${\bf p}$, taking into account the Green function structure of Eq.(\ref{G1}), one discovers that several components of $S_{(a)}^{\rho\lambda}$ given in Eq.(\ref{WFa1}) vanish.  A similar analysis can be done for the weight factors $S_{(b)}^{\rho\lambda}$ of Eq.(\ref{WFb}) when considering integration over the directions of ${\bf p}$ and ${\bf p'}$.
As a result, the only non vanishing weight factors are
\begin{equation}
\begin{split}
\label{WFaCC}
S^{00}_{(a)}&=-e^{2}(2\pi N_{0}\tau)2D\tau\Big(1-\frac{\tau}{\tau_{DP}}\Big)\,,\\
S^{11}_{(a)}&=e^{2}(2\pi N_{0}\tau)2D\tau\frac{\tau}{4\tau_{DP}}\,,\\
 S^{22}_{(a)}&=e^{2}(2\pi N_{0}\tau)2D\tau\frac{3\tau}{4\tau_{DP}}\,
 \end{split}
\end{equation}
and
\begin{equation}
\begin{split}
\label{WFbCC}
\text{Re}S^{00}_{(b)}&=e^{2}\frac{2\pi N_{0}\tau}{4}D\tau\frac{x^{2}}{(1+x^{2})^{2}}\,,\\
\text{Re}S^{22}_{(b)}&=e^{2}\frac{2\pi N_{0}\tau}{4}D\tau\frac{x^{2}}{(1+x^{2})^{2}}\,,
 \end{split}
\end{equation}
where we have neglected terms beyond the leading order in $1/(\epsilon_{F}\tau )$ and the first order in $\alpha/v_{F}$.
By inserting the weight factors (\ref{WFaCC}) in Eq. (\ref{a.1}), the quantum correction, given by the diagram of Fig.\ref{bubble}, is
\begin{equation}
\begin{split}
\delta\sigma_{(a)}
&=\frac{\sigma_{0}}{2\pi\epsilon_{F}\tau}\Bigg[\log\frac{L}{l}-\frac{1+x^2/2}{2(1+x^2)}\log\Bigg(\frac{1+\frac{2\tau}{\tau_{DP}}}{l^2/L^2+\frac{2\tau}{\tau_{DP}}}\Bigg)-\\
&-\frac{2+x^2/2}{4(1+x^2)}\log\Bigg(\frac{1-\frac{2\tau}{\tau_{DP}}+\frac{8\tau^{2}}{\tau^{2}_{DP}}}{l^{4}/L^4-\frac{2\tau}{\tau_{DP}}l^2/L^2+\frac{8\tau^{2}}{\tau^{2}_{DP}}}\Bigg)\Bigg]\end{split}\label{QCCa}.
\end{equation}
 The quantum correction, given by the diagrams of Figs. \ref{up} and \ref{down}, is
\begin{eqnarray}
\delta\sigma_{(b)}+\delta\sigma_{(c)}&=&\frac{\sigma_{0}}{2\pi\epsilon_{F}\tau}\frac{x^{2}}{4(1+x^{2})^{2}}\Bigg[\frac{1}{2}\log\Bigg(\frac{1+\frac{2\tau}{\tau_{DP}}}{l^2/L^2+\frac{2\tau}{\tau_{DP}}}\Bigg)  \nonumber \\
&+& \frac{3}{4}\log\Bigg(\frac{1-\frac{2\tau}{\tau_{DP}}+\frac{8\tau^{2}}{\tau^{2}_{DP}}}{l^{4}/L^4-\frac{2\tau}{\tau_{DP}}l^2/L^2+\frac{8\tau^{2}}{\tau^{2}_{DP}}}\Bigg)\Bigg]\,.
\label{QCCb}
\end{eqnarray}
To our knowledge the expression for diagrams of Figs. \ref{up} and \ref{down} has not been given before.
 In agreement with \cite{Edelstein95} the contribution of the two diagrams \ref{up} and \ref{down} vanishes in the absence of Rashba SOC and does not  exhibit scaling behavior for $x\neq0$. So it can be ignored and the quantum correction to the electrical conductivity is given by $\delta\sigma_{(a)}$ \cite{Iordanskii_1994,Skvortsov_98,Lyanda98,Araki_2014} .
 \begin{figure}
\begin{center}
\includegraphics[width=0.85\textwidth]{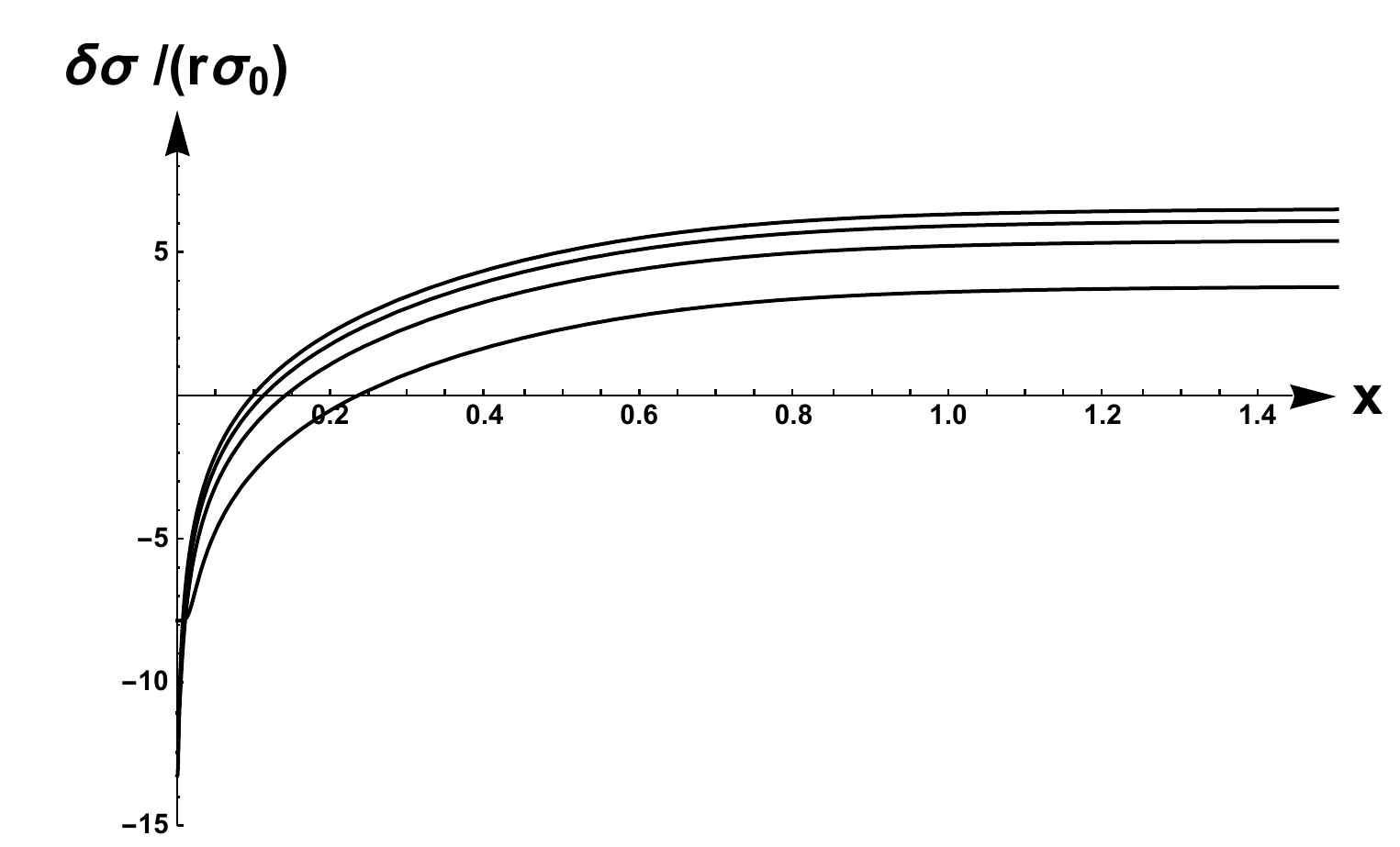}
\caption{The quantum correction to the electrical conductivity as a function of the variable $x=2\alpha p_{F}\tau$ for four different values of the ratio $L/l=50,250,500,750$ from bottom to top.
In the $y$ axis we set $\delta\sigma/(r\sigma_{0} )$, where $r=1/(2\pi\epsilon_{F}\tau )$.}
\label{Scaling}
\end{center}
\end{figure}

The above correction (\ref{QCCa}) can be interpreted as a sum of a localizing contribution from the triplet sector and an antilocalizing contribution from the singlet channel.
Therefore, for $l/L\ll x$, the triplet modes acquire a gap that reduces their contribution and the total correction changes its sign and becomes antilocalizing at a certain $x^{*}$.
The critical value $x^{*}$, which defines the crossover point between WL and WAL, depends on the ratio $L/l$.
In particular, as the ratio $L/l$ gets larger, the value $x^{*}$ decreases\cite{Skvortsov_98,Araki_2014}.
This behavior is evident by looking at Fig.  \ref{Scaling}; for $x=0$ the quantum corrections are localizing, whereas for $x>x^{*}$ the quantum correction tends to
\begin{equation}
\label{QCECC1}
\delta\sigma= \frac{\sigma_{0}}{2\pi\epsilon_{F}\tau}\log\frac{L}{l}\,,
\end{equation}
which implies the enhancement of the electrical conductivity and the diffusion coefficient.

\section{The weak-localization corrections to the EC and SHC}

The   equation (\ref{QCDC}) for the quantum correction to the electrical conductivity can easily be generalized to the spin transport coefficients by substituting in the weight factors (\ref{WFa1}) and (\ref{WFb}) the left charge current vertex with the renormalized spin density vertex  or the renormalized spin current vertex.  The correction to the Edelstein conductivity then reads
\begin{equation}
\label{QCEC}
\delta\sigma^{EC}=\frac{1}{2\pi}\sum_{\rho\mu\nu\lambda}\text{Tr}(\sigma^{\mu}\sigma^{\lambda}\sigma^{\nu}\sigma^{\rho})(W^{\rho\lambda}_{(a)}+2\text{Re}W^{\rho\lambda}_{(b)})C^{\mu\nu}\,,
\end{equation}
with weight factors
\begin{equation}
\label{WFECa}
W^{\rho\lambda}_{(a)}=-\sum_{{\bf p}}\frac{1}{2}\text{Tr}\Big[G^{A}_{{\bf p}}\frac{\tau_{DP}}{2\tau}\sigma^{y}G^{R}_{{\bf p}}\sigma^{\rho}\Big]\frac{1}{2}\text{Tr}\Big[G^{R}_{-{\bf p}}(-e)\frac{p_{x}}{m}G^{A}_{-{\bf p}}\sigma^{\lambda}\Big]\,,
\end{equation}
and
\begin{equation}
\label{WFECb}
W^{\rho\lambda}_{(b)}=-\frac{1}{2\pi N_{0}\tau}\sum_{{\bf p},{\bf p}^{\prime}}\frac{1}{2}\text{Tr}\Big[G^{A}_{{\bf p}}\frac{\tau_{DP}}{2\tau}\sigma^{y}G^{R}_{{\bf p}} G^{R}_{{\bf p}^{\prime}}\sigma^{\rho}\Big]\frac{1}{2}\text{Tr}\Big[G^{R}_{-{\bf p}} G^{R}_{-{\bf p}^{\prime}} (-e)\frac{p^{\prime}_{x}}{m}G^{A}_{-{\bf p}^{\prime}}\sigma^{\lambda}\Big]\,.
\end{equation}
Notice that in the  right vertex it appears the renormalized charge current vertex $J_x$, whereas in the leading order expression (\ref{8}) the bare charge vertex $j_x$ appears.
The non-vanishing weight factors are:
\begin{equation}
\begin{split}
\label{WeightfactorsEc}
\overline{W}^{00}&=e\frac{\tau_{DP}}{2\tau}\frac{2\pi N_{0}\tau}{4}\alpha\tau^{2}\frac{5x^2+6x^4+2x^6}{(1+x^2)^3}\,,\\
\overline{W}^{11}&=-e\frac{\tau_{DP}}{2\tau}\frac{2\pi N_{0}\tau}{4}\alpha\tau^{2}\frac{-x^{2}+x^4+x^6}{(1+x^{2})^{3}}\,,\\
\overline{W}^{22}&=-e\frac{\tau_{DP}}{2\tau}\frac{2\pi N_{0}\tau}{4}\alpha\tau^{2}\frac{x^{2}+3x^{4}+x^6}{(1+x^{2})^{3}}\,,\\
\overline{W}^{33}&=e\frac{\tau_{DP}}{2\tau}\frac{2\pi N_{0}\tau}{4}\alpha\tau^{2}\frac{x^{2}(1+2x^2)}{(1+x^{2})^{3}}\,,
\end{split}
\end{equation}
and $\overline{W}^{\rho\lambda}=0$ if $\rho\neq\lambda$.
In the above equation we have defined $\overline{W}^{\rho\lambda}=W^{\rho\lambda}_{(a)}+2\text{Re}W^{\rho\lambda}_{(b)}$.
Since the weight factors are diagonal in the indices $\lambda$ and $\rho$, only the diagonal elements $C^{\mu\mu}$ contribute to the correction to the Edelstein conductivity Eq.(\ref{QCEC}).
If we take into account only the singlet channel, which is the only mode with scaling behavior for $x\neq0$, then all the diagonal elements $C^{\mu\mu}$ are related:
$C^{ii}=-C^{00}$ with $i=1,2,3$ and
\begin{equation}
\label{singletpart}
C^{00}=\frac{1}{4\tau^2}\frac{1}{2\pi \epsilon_F\tau}\log \frac{L}{l}.
\end{equation}
Then, by performing the trace over the product of Pauli matrices in Eq.(\ref{QCEC}), one obtains
\begin{equation}
\label{QCEC1}
\delta\sigma^{EC}=\frac{1}{2\pi} 4 (-\overline{W}^{00}+
\overline{W}^{11}+\overline{W}^{22}+\overline{W}^{33}) C^{00},
\end{equation}
from which
the quantum correction to the Edelstein conductivity reads
 \begin{equation}
\label{WLEdeltsein}
\delta\sigma^{EC}=\frac{\sigma^{EC}_{0}}{2\pi\epsilon_{F}\tau}\log\frac{L}{l}\,.
\end{equation}
The above contribution increases the leading order value
\begin{equation}
\label{WLEdeltsein1}
\sigma^{EC}=-e\alpha N_{0}\tau \Big(1+\frac{1}{2\pi\epsilon_{F}\tau}\log\frac{L}{l}\Big)\,,
\end{equation}
which shows that the logarithmic correction to the Edelstein conductivity can be absorbed in terms of the renormalization of the elastic scattering time.
Thus, Rashba SOC provides the spin relaxation mechanism, which causes the change of sign of the weak localization corrections and   does not introduce a new scaling parameter.

For the evaluation of the spin Hall conductivity, we rely on the decomposition shown in the first line of Eq.(\ref{SHC5}),
where here we replace the bare charge current verices $j_x$ with the renormalized one $J_x$ and insert a series of maximally crossed impurity lines between the two vertices.
Hence, the resulting expression will have a Cooperon connected to a Hikami box with modified vertices.
The resulting expression reads
\begin{equation}
\label{QCSHc}
\delta\sigma^{SHC}=\frac{1}{e}\sigma_0\delta\gamma_{SH}+\frac{v_{F}}{2}\frac{x}{1+x^{2}}\delta\sigma^{EC}\,,
\end{equation}
where
\begin{equation}
\label{QCSHc1}
\frac{1}{e}\sigma_0\delta\gamma_{SH}=\frac{1}{2\pi}\sum_{\rho\mu\nu\lambda}\text{Tr}(\sigma^{\mu}\sigma^{\lambda}\sigma^{\nu}\sigma^{\rho})(L^{\rho\lambda}_{(a)}+2\text{Re}L^{\rho\lambda}_{(b)})C^{\mu\nu}\,.
\end{equation}
Notice that the response function indicated by $\sigma_0 \delta\gamma_{SH}$ contains the renormalized charge current vertex $J_x$ in contrast with $\sigma^{SHC}_{drift}$ which has the bare charge current vertex. This is the reason why we only find the term $\sigma_0 \delta\gamma_{SH}$ and we don't find $\delta\sigma_0 \gamma_{SH}$ in Eq.(\ref{QCSHc}).
The response function $\sigma_0 \delta\gamma_{SH}$ describes the influence of the WL corrections on the spin Hall angle.

The weight factors are
\begin{equation}
\label{WFSHCa}
L^{\rho\lambda}_{(a)}=-\sum_{{\bf p}}\frac{1}{2}\text{Tr}\Big[G^{A}_{{\bf p}}\frac{p_{y}}{m}\frac{\sigma^{z}}{2}G^{R}_{{\bf p}}\sigma^{\rho}\Big]\frac{1}{2}\text{Tr}\Big[G^{R}_{-{\bf p}}(-e)\frac{p_{x}}{m}G^{A}_{-{\bf p}}\sigma^{\lambda}\Big]\,,
\end{equation}
and
\begin{equation}
\label{WFSHCb}
L^{\rho\lambda}_{(b)}=-\frac{1}{2\pi N_{0}\tau}\sum_{{\bf p},{\bf p}^{\prime}}\frac{1}{2}\text{Tr}\Big[G^{A}_{{\bf p}}\frac{p_{y}}{m}\frac{\sigma^{z}}{2}G^{R}_{{\bf p}} G^{R}_{{\bf p}^{\prime}}\sigma^{\rho}\Big]\frac{1}{2}\text{Tr}\Big[G^{R}_{-{\bf p}} G^{R}_{-{\bf p}^{\prime}} (-e)\frac{p^{\prime}_{x}}{m}G^{A}_{-{\bf p}^{\prime}}\sigma^{\lambda}\Big]\,.
\end{equation}
The only non-vanishing weight factors are:
\begin{equation}
\begin{split}
\label{WeightfactorsShc}
\bar{L}^{00}&=-e\frac{\tau}{m}\frac{2\pi N_{0}\tau}{16}\frac{4 x^{2} +4x^{4}+x^{6}}{(1+x^{2})^{3}}\,,\\
\bar{L}^{11}&=-e\frac{\tau}{m}\frac{2\pi N_{0}\tau}{16}\frac{2 x^{2}-x^6}{(1+x^{2})^{3}}\,,\\
\bar{L}^{22}&= e\frac{\tau}{m}\frac{2\pi N_{0}\tau}{16}\frac{2 x^{2} +4x^{4}+x^{6}}{(1+x^{2})^{3}}\,,\\
\bar{L}^{33}&=e\frac{\tau}{m}\frac{2\pi N_{0}\tau}{16}\frac{x^{6}}{(1+x^{2})^{3}}\,,
\end{split}
\end{equation}
and $\bar{L}^{\rho\lambda}=0$ if $\rho\neq\lambda$.
In the above equation we have defined $\bar{L}^{\rho\lambda}=2\text{Re}L^{\rho\lambda}_{(b)}$, since the contribution of the first diagram of the Hikami boxes is equal to zero, $\sigma_0\delta\gamma_{SH,(a)}=0$.
Finally, by replacing the weight factors $\overline{W}^{\lambda\lambda}$ with the $\bar L^{\lambda\lambda}$ in Eq.(\ref{QCEC1}), the quantum correction to the {\it drift} component reads
\begin{equation}
\label{WLbareSHc}
\frac{1}{e} \sigma_0\delta\gamma_{SH}=\frac{e}{8\pi}\frac{x^2 (1+x^{2})^{2}}{(1+x^2)^3}\frac{1}{2\pi m D}\log\frac{L}{l}=\sigma_{drift}^{SHC}\frac{1}{2\pi\epsilon_{F}\tau}\log\frac{L}{l}\,,
\end{equation}
where we have only taken the logarithimic contribution given by the singlet channel.
It is then evident that the insertion of the results (\ref{WLbareSHc}) and (\ref{WLEdeltsein}) into (\ref{QCSHc}) gives the vanishing of the weak localization correction to the spin Hall conductivity
\begin{equation}
\label{DMCWL}
\delta\sigma^{SHC}=\Big(\sigma^{SHC}_{drift}+\frac{v_{F}}{2}\frac{x}{1+x^{2}}\sigma^{EC}_{0}\Big)\frac{1}{2\pi \epsilon_{F}\tau}\log\frac{L}{l}=0\,
\end{equation}
as required by the relation (\ref{Di2}) in the static limit.

\section{Discussion}

In this Section we develop an interpretation of the structure of the weak localization results of the previous Section. By considering Eq.(\ref{discussion_1}) connecting spin current and spin density, we assume for the WL corrections the structure
\begin{equation}
\label{interpretation_1}
\delta \langle\langle j^z_y\rangle\rangle={2m\alpha} (\delta D)\langle \langle s^y\rangle\rangle +
{2m\alpha} D (\delta \langle\langle s^y\rangle\rangle) +\frac{(\delta \gamma_{SH})}{e}\langle\langle j_x\rangle\rangle+\frac{\gamma_{SH}}{e}(\delta \langle\langle j_x\rangle\rangle),
\end{equation}
where the variation has been applied to both the coefficients connecting the averaged values of the observables and to the average of the observables themselves.
In the limit $x\ll 1$, by noticing that $v_F x/2=2m\alpha D$,
we identify  the second term in the right hand side of the above equation with the second term in the right hand side of Eq.(\ref{DMCWL}).  In the same limit, i.e. keeping terms up to order $\alpha^2$, we see that the first and last term of Eq.(\ref{interpretation_1}) must necessarily cancel among themselves. Hence, one can identify the third term of
Eq.(\ref{interpretation_1}) with the first term of Eq.(\ref{DMCWL}), thus making clear the reason for introducing the response function $\sigma_0\delta \gamma_{SH}$ describing the quantum correction to the spin Hall angle.

The direct measurement of the EC or its inverse is possible\cite{RojasSanchez2013,ganichev2002} and 
a possible test of the theory presented here would be the study of the EC as a function of an applied magnetic field perpendicular to the 2DEG. 
On  the other hand, the SHC  vanishes in the Rashba disordered 2DEG. 
It is then important to extend this analysis to a case where the SHC is finite, as for instance when also extrinsic SOC from impurities is present\cite{Raimondi09}. This however is beyond the scope of the present work.

{\bf Aknowledgements} R.R. thanks Valentina Brosco and 
Lara Benfatto for discussions.

\appendix

\section{The Cooperon in the presence of Rashba spin-orbit coupling}
\label{CRSOC}
To describe the sum of maximally crossed diagrams, we have introduced the so-called Cooperon, corresponding to the diagram depicted in Fig. \ref{fig:Cooperon}.
 In the main text we have derived an approximate solution to the Cooperon equation, by using an argument proposed by Kettemann and Wenk \cite{Wenk_10}.
 In this appendix we want to obtain the exact derivation of the Cooperon.

As  can be understood by looking at Fig. \ref{fig:Cooperon}, the equation for the Cooperon can be written as
\begin{equation}
\label{Directproduct}
C^{-1}({\bf Q})=2\pi N_{0}\tau\big(1-\sum_{\mu\nu}L^{\mu\nu}\,I^{\mu\nu}({\bf Q})\big)\,,
\end{equation}
where $L^{\mu\nu}=\sigma^{\mu}\otimes\sigma^{\nu}$ and
 \begin{equation}
\label{I}
I^{\mu\nu}({\bf Q})=\frac{1}{2\pi N_{0}\tau}\sum_{{\bf p}}\left(\begin{array}{c}G^{R}_{0} \\ie^{-i\phi}\Delta G^{R} \\-ie^{i\phi}\Delta G^{R} \\0\end{array}\right)^{\mu}_{{\bf p}}\left(\begin{array}{cccc}G^{A}_{0}, &ie^{-i\phi}\Delta G^{A}, &-ie^{i\phi}\Delta G^{A}, &0\end{array}\,\right)^{\nu}_{{\bf Q-p}}\,,
\end{equation}
where $\mu=0,+,-,3$, $G_{0}=(G_{+}+G_{-})/2$, $\Delta G=(G_{+}-G_{-})/2$ and $\phi=\arctan(p_{y}/p_{x})$.
%In the above equation (\ref{Cooperon}) we have neglected a small non-divergent contribution.
 Since we expect the Cooperon to be large in vicinity of backscattering ${\bf p}=-{\bf p}^{\prime}$, we set ${\bf Q}=Q(\cos\theta,\sin\theta)$ and expand $\hat{I}({\bf Q})$ in powers of Q up to the quadratic order:
  \begin{equation}
\label{expansionQ}
I^{\mu\nu}({\bf Q})=I^{\mu\nu}_{(0)}+I^{\mu\nu}_{(1)}(\theta)Q+I^{\mu\nu}_{(2)}(\theta)Q^{2}\,,
\end{equation}
where the expansion coefficients are given by
\begin{equation}
\begin{split}
\label{GE}
I^{\mu\nu}_{(0)}&=\frac{1}{2\pi N_{0}\tau}\sum_{{\bf p}}G^{R,\mu}({\bf p})G^{A,\nu}(-{\bf p})\,,\\
I^{\mu\nu}_{(1)}&=\frac{1}{2\pi N_{0}\tau}\sum_{{\bf p}}G^{R,\mu}({\bf p})\frac{\partial G^{A,\nu}({\bf Q}-{\bf p})}{\partial Q}\Big|_{Q=0}\,,\\
I^{\mu\nu}_{(2)}&=\frac{1}{2\pi N_{0}\tau}\sum_{{\bf p}}G^{R,\mu}({\bf p})\frac{\partial^{2} G^{A,\nu}({\bf Q}-{\bf p})}{\partial Q^{2}}\Big|_{Q=0}\,.
\end{split}
\end{equation}
%Finally, once we evaluate the integrals (\ref{GE}), in the direct product basis the Cooperon will be

\paragraph{Zeroth order: $I^{\mu\nu}_{(0)}$}
By symmetry due to the integration over the angle of ${\bf p}$, there are only three integrals different from zero $I^{00}_{(0)}$, $I^{11}_{(0)}$ and $I^{22}_{(0)}$.
In particular one has
\begin{equation}
\begin{split}
\label{Orderzero}
I^{00}_{(0)}=1-\frac{x^{2}}{2(1+x^{2})}=1-\frac{\tau}{\tau_{DP}}\,,\\
I^{+-}_{(0)}=I^{-+}_{(0)}=-\frac{x^{2}}{2(1+x^{2})}=-\frac{\tau}{\tau_{DP}}
\end{split}
\end{equation}
where we have introduced the D'yakonov-Perel spin relaxation time $\tau_{DP}$.
In the direct product basis $\big\{
|\uparrow\uparrow\rangle, |\uparrow\downarrow\rangle, |\downarrow\uparrow\rangle, |\downarrow\downarrow\rangle\big\}$ the ${\bf Q}=0$ expression of the Cooperon is:
 \begin{equation}
\label{Q=0}
C^{-1}_{(0)}=2\pi N_{0}\tau(1-I_{(0)})=2\pi N_{0}\tau\left(\begin{array}{cccc}\tau/\tau_{DP}& 0 & 0 & 0 \\0 & \tau/\tau_{DP} & \tau/\tau_{DP} & 0 \\0 & \tau/\tau_{DP} & \tau/\tau_{DP} & 0 \\0 & 0 & 0 & \tau/\tau_{DP}\end{array}\right)\,.
\end{equation}

%The eigenvalues of $C^{-1}$ are the triplet and singlet states.
%This is evident from the fact that first and last states of the basis are the triplet states with total spin projection $\pm1$.
%The diagonalization of the central two by two block gives as eigenstates the triplet state with spin 0 and eigenvalue $2\tau/\tau_{DP}$ and the singlet state with eigenvalue 0.

\paragraph{First order: $I^{\mu\nu}_{(1)}$}
At linear order in the $Q$ we have only four contributions:
\begin{equation}
\begin{split}
\label{Orderzero2}
&I^{+0}_{(1)}(\theta)=-\frac{i\,e^{-i\theta}}{4}J\,,\quad I^{0+}_{(1)}(\theta)=-\frac{i\,e^{-i\theta}}{4}J^{*}\,,\\
&I^{-0}_{(1)}(\theta)=\frac{i\,e^{i\theta}}{4}J\,,\quad I^{0-}_{(1)}(\theta)=\frac{i\,e^{i\theta}}{4}J^{*}\,,
\end{split}
\end{equation}
each one is expressed in terms of the integral
\begin{equation}
\label{basicintegral}
J\simeq\frac{1}{2\pi N_{0}\tau}\sum_{{\bf p}}\frac{p}{m}\Delta G^{R}_{{\bf p}}({G^{A}_{{\bf p}+}}^{2}+{G^{A}_{{\bf p}-}}^{2})\,.
\end{equation}
where in the $Q$ expansion we have taken the leading order in $\alpha/v_{F}$ and $1/\epsilon_{F}\tau$. According to the previous approximation we obtain
\begin{equation}
\label{J}
J=2v_{F}\tau\frac{x}{(1+x^{2})^{2}}\,.
\end{equation}

In the direct product basis the linear term in $Q$ has the following expression
\begin{equation}
\label{Linear}
C^{-1}_{(1)}(\theta)=-2\pi N_{0}\tau I_{(1)}=2\pi N_{0} \tau\frac{\tau v_{F} x}{2(1+x^{2})^{2}} \left(\begin{array}{cccc}0& ie^{-i\theta} & ie^{-i\theta} & 0 \\-ie^{i\theta}& 0 & 0 & ie^{-i\theta} \\-ie^{i\theta} & 0 & 0& ie^{-i\theta} \\0 & -ie^{i\theta}& -ie^{i\theta}&0\end{array}\right)\,.
\end{equation}

\paragraph{Second order: $I^{\mu\nu}_{(2)}$}
The second order contribution is given by five integrals $I^{00}_{(2)}$, $I^{++}_{(2)}$, $I^{--}_{(2)}$, $I^{+-}_{(2)}$ and $I^{-+}_{(2)}$.
 At leading order in $\alpha/v_{F}$ and $1/\epsilon_{F}\tau$, these coefficients read
 \begin{equation}
\begin{split}
\label{order2}
I^{00}_{(2)}&=-D\tau\Big[1-\frac{x^{2}(6+3x^{2}+x^{4})}{2(1+x^{2})^{3}}\Big]\,,\\
I^{++}_{(2)}(\theta)&=-e^{-2i\theta}D\tau\frac{x^{2}(6+3x^{2}+x^{4})}{4(1+x^{2})^{3}}=\big(I^{--}_{(2)}(\theta)\big)^{*}\,,\\
I^{+-}_{(2)}&=-D\tau \frac{x^{2}(6+3x^{2}+x^{4})}{2(1+x^{2})^{3}}=I^{-+}_{(2)}\,,
\end{split}
\end{equation}
where $D=v^{2}_{F}\tau/2$. In the direct product basis $Q^{2}$ is multiplied for
\begin{equation}
\label{onematrix}
C^{-1}_{(2)}(\theta)=2\pi\tau N_{0}D\tau \left(\begin{array}{cccc}1-h(x)&0 & 0& h(x)e^{-2i\theta}/2 \\0& 1-h(x) & -h(x) & 0 \\0 & -h(x) & 1-h(x)& 0 \\ h(x)e^{2i\theta}/2  &0& 0& 1-h(x)\end{array}\right)\,,
\end{equation}
where $h(x)=x^{2}(6+3x^{2}+x^{4})/2(1+x^{2})^{3}$.

Thus, in the triplet singlet basis the Cooperon reads
\begin{equation}
\begin{split}
\label{CTS}
C^{-1}({\bf Q})=&2\pi N_{0}\tau\Big[D\tau Q^{2}+(\hat{{\bf S}}^{2}-\hat{S}^{2}_{z})\Big(\frac{\tau}{\tau_{DP}}-D\tau Q^{2}\frac{h(x)}{2}\Big)-D\tau h(x)({{\bf Q}}\times\hat{{\bf S}})^{2}-\\
&-v_{F}\tau \frac{x}{(1+x^{2})^{2}}({\bf Q}\times\hat{{\bf S}})\cdot\hat{z}+h(x) D\tau Q^{2} \hat{S}^{2}_{z}\Big]
\end{split}
\end{equation}
where $\hat{z}$ is the versor perpendicular to the plane of the 2DEG and  $\hat{{\bf S}}$ is the total spin operator for a pair of electrons.
The above result agrees with \cite{Skvortsov_98,Araki_2014}. However, the most important effect due to the SOC is the appearance of the spin relaxation time, which cut off the triplet channels.
% and in the dirty regime, $x\ll1$, it becomes that
This permit us to use the diffusive limit Cooperon (\ref{KW1}), proposed in \cite{Lyanda98,Wenk_10}.

\section{Integrals of Green functions products}

To perform the calculations of the weak localization corrections of the transport coefficients we encounter the following kind of integral
\begin{equation}
\label{Genericintegral}
\sum_{{\bf p}}p^{l}{G^{R}_{\pm}}^{n}({\bf p}){G^{A}_{\pm}}^{m}({\bf p})
\end{equation}
 where $\pm$ are the band indices, $n$, $m$ and $l$ three integer numbers.
 When the chemical potential is the largest energy scale, we are able to transform the sum over momentum as
 \begin{equation}
\label{repleceament}
\sum_{{\bf p}}(\cdots)=N_{0}\int_{0}^{2\pi}\frac{d\phi}{2\pi}\int_{-\infty}^{\infty}d\xi(\cdots)
\end{equation}
where $\xi=p^{2}/2m-\mu$. Then the integral over $\xi$ is carried by evaluating the residues to the first order in $\alpha/v_{F}$ and to the zero order in $1/(\epsilon_{F}\tau )$.
However, this requires some care when the integral involves the two Fermi surfaces and the parameter $x=2\alpha p_F \tau$ appears. Indeed, since $x=(\alpha /v_F) (4\epsilon_F \tau)$,  then when $x$ is present one must keep also terms of order $1/(\epsilon_F\tau)$, which eventually yield terms of order $\alpha/v_F$ when multiplied by $x$.

Below we report the results:
\begin{equation}
\begin{split}
\label{Integrals1}
&\sum_{{\bf p}}p\,{G^{R}_{\pm}}^{2}({\bf p}){G^{A}_{\pm}}^{2}({\bf p})=4\pi N_{0}\tau^{3} p_{F}\Big(1\mp\frac{2\alpha}{v_{F}}\Big)\,,\\
&\sum_{{\bf p}}p\,G^{R}_{+}({\bf p})G^{R}_{-}({\bf p}){G^{A}_{\pm}}^{2}({\bf p})=2\pi N_{0}\tau^{3}p_{F}\frac{2\mp i\,x}{(1\mp ix)^2}\Big(1\mp\frac{\alpha}{v_{F}}\Big)\,,
\end{split}
\end{equation}
 and those that appear in the Hikami boxes corrected by the impurity line
\begin{equation}
\begin{split}
\label{Integrals2}
&\sum_{{\bf p}}p\,{G^{R}_{\pm}}^{2}({\bf p}){G^{A}_{\pm}}({\bf p})=-2i\pi N_{0} \tau^{2} p_{F}\Big(1\mp\frac{2\alpha}{v_{F}}\Big)\,,\\
&\sum_{{\bf p}}p\,G^{R}_{+}({\bf p})G^{R}_{-}({\bf p})G^{A}_{\pm}({\bf p})=-2i\pi N_{0} \tau^{2} p_{F}\frac{1\mp\frac{\alpha}{v_{F}}-\frac{i}{4\tau\epsilon_F}}{1\mp ix}\,,\\
&\sum_{{\bf p}}p\,{G^{R}_{\pm}}^{2}({\bf p}){G^{A}_{\mp}}({\bf p})=-2i\pi N_{0} \tau^{2} p_{F}\frac{1-\frac{i}{4\tau\epsilon_F}}{(1\pm ix)^{2}}\,,\\
&\sum_{{\bf p}}\,{G^{R}_{\pm}}^{2}({\bf p}){G^{A}_{\pm}}({\bf p})=-2i\pi N_{0}\tau^{2}\Big(1\mp\frac{\alpha}{v_{F}}\Big)\,,\\
&\sum_{{\bf p}}\,G^{R}_{+}({\bf p})G^{R}_{-}({\bf p})G^{A}_{\pm}({\bf p})=-2i\pi N_{0}\tau^{2}\frac{1}{1\mp ix}\,,\\
&\sum_{{\bf p}}\,{G^{R}_{\pm}}^{2}({\bf p}){G^{A}_{\mp}}({\bf p})=-2i\pi N_{0}\tau^{2}\frac{1\mp\frac{\alpha}{v_{F}}}{(1\pm ix)^2}\,.
\end{split}
\end{equation}
Finally, we report the integrals that occur in the weak localization correction to the electrical conductivity
\begin{equation}
\begin{split}
\label{Integrals3}
&\sum_{{\bf p}}p^{2}\,{G^{R}_{\pm}}^{2}({\bf p}){G^{A}_{\pm}}^{2}({\bf p})=4\pi N_{0}\tau^{3} p^{2}_{F}\Big(1\mp\frac{3\alpha}{v_{F}}\Big)\,,\\
&\sum_{{\bf p}}p^{2}\,G^{R}_{+}({\bf p})G^{R}_{-}({\bf p})G^{A}_{+}({\bf p})G^{A}_{-}({\bf p})=4\pi N_{0}\tau^{3}p^{2}_{F}\frac{1}{1+x^2}\,.
\end{split}
\end{equation}

\bibliography{ref}

\end{document}